\documentclass[10pt]{article}
\usepackage{graphicx}
\usepackage{amssymb, amsmath, amsthm, amsfonts, bm, mathrsfs, wasysym, txfonts, bbm, enumerate, color}

\begin{document}
\title{Surface Configuration in $R+\mu^4/R$ Gravity}

\author{Mohsen Fathi\thanks{email: mohsen.fathi@gmail.com} ,  Morteza Mohseni\thanks{email: m-mohseni@pnu.ac.ir}}

\maketitle \centerline{\it Department of Physics, Payame Noor
University (PNU), PO BOX 19395-3697 Tehran, Iran}

\begin{abstract}
We investigate the conditions on the
additional constant $\mu$ in the so-called $R+\mu^4/R$ theory of
gravity, due to existence of different kinds of space-like
surfaces in both weak field and strong field limits, and their
possible correspondence to black hole event horizons. Adopting a Schwarzschild limit, 
we probe the behaviour of $\mu$ in different contexts of radial and radial-rotational
congruence of null geodesics. We show that these cases serve as correspondents to 
black hole horizons in some peculiar cases of study.
 \\\\
{\textit{keywords}}: $f(R)$ gravity, Black hole horizons, Null
geodesic flows, Raychaudhuri equation
\end{abstract}

\section{Introduction}
The mathematical features of a black hole, depending on peculiar
singularities in solutions of Einstein field equations, became
more interesting after the concept of an Event Horizon was
introduced by Finkelstein \cite{Finkelstein1958}. This was when he
noted a null 2-dimensional hypersurface\footnote{Usually horizons are considered to be 3-dimensional hypersurfaces. The surfaces we discuss here are those which are
obtained by foliating such three-dimensional horizons by means of 2-surfaces.} in Schwarzschild spacetime,
through which light rays can pass, but they will be trapped
beyond. Even though this is not the only way of defining a
horizon, however it appeared to be of the most interest. Therefore
since no one can interact with what is beyond a horizon, the only
way that one can investigate the physics of a black hole, is to
examine its horizon or strictly speaking, the space-like surface
which conceals the black hole. Mostly, black holes are
mathematically extrapolated from solutions of general relativity.
On the other hand, a great deal of physicists' attention have been
devoted to alternatives to general relativity, specially after
discovery of accelerated expansion of universe
in the late of 90's, according to the observed anomalous redshift of SNIa \cite{Riess1998, Perlmutter1999}.
These observations somehow led to the appearance of dark
energy concepts. Moreover, the observation of galactic flat rotation 
curves, led to the advent of dark matter scenario \cite{Persic1996}.

However some believe that our inability to
properly explain this accelerated expansion, stems from our
misinterpretations of gravitation. So it seems that proposing a
viable alternative to general relativity would be of benefit.
However as it was stated above, the important concept of black
holes should be valid in a gravitational theory. Therefore, there
have been so much endeavors to obtain same results in modified
theories \cite{Cruz-Dombriz2009}.

One of the most essential alternatives to general relativity which
this article is mostly devoted to a peculiar class of it, is
$f(R)$ theories (introduced in the next section) and black hole
physics has been also discussed in this theory (e.g. see
\cite{Cruz-Dombriz2009b, Myung2011, Nzioki2014}). Therefore it
seems plausible to look for the behaviour of the so-called
space-like surfaces on possible black-holes. In this paper, we are
about to take care about this problem, by considering a peculiar
class of $f(R)$ theories, namely the $R+\mu^4/R$ theory of
gravity.

The paper is organized as follows: in section 2 we bring a
constructive review on the $f(R)$ gravity and its special case,
$R+\mu^4/R$ theory. In section 3, we use a weak field static
solution of the theory to investigate the behaviour of null-like
geodesic congruence which are passing the surfaces and discuss
the possible types. Same procedure is exploited for the strong
filed limit of the theory in section 4. We summarized in section
5.

\section{$R+\mu^4/R$ Theory of Gravity} 
In the metric formalism, the general $f(R)$ action
is written as \cite{Sotiriou2010}
\begin{equation}\label{1}
   S_{f(R)}=\frac{1}{2\kappa^2}\int\textrm{d}^4x\sqrt{-g}\,\,f(R)+\int\textrm{d}^4x\,\,\mathcal{L}_m\left(g_{ab,\Psi_m}\right),
\end{equation}
where $f(R)$ is an arbitrary (analytic) function of the Ricci scalar curvature\footnote{In the metric formalism, $R$ is obtained from the standard metric, 
while in Palatini approach, this scalar and variations are in terms of an independent connection \cite{Sotiriou2010}.} of spacetime \cite{Capozziello2008}
and $\mathcal{L}_m$ is the matter Lagrangian for perfect fluids,
depending on the spacetime metric $g_{ab}$ and the matter/energy
fields $\Psi_m$. There have been also studies on the cases in
which the Ricci scalar and matter are coupled \cite{Bertolami2007,
Bertolami2008, Faraoni2009}. The standard metric variation with
respect to $g_{ab}$ provides the field equations
\begin{equation}\label{2}
    f'(R)R_{ab}-\frac{1}{2} f(R)
    g_{ab}-\nabla_a\nabla_b f'(R)+g_{ab}\Box
    f'(R)=\kappa^2 T_{ab},
\end{equation}
with $f'(R)=\frac{\textrm{d}f}{\textrm{d}R}$ and
$T_{ab}=-\frac{2}{\sqrt{-g}}\frac{\delta\mathcal{L}_m}{\delta
g^{ab}}$. Evidently, for $f(R)=R$, the field equations (\ref{2})
will regain the Einstein field equations of general relativity.
However, the most intuitive alternative conjecture is
$f(R)=R+\alpha R^2$, proposed by Starobinsky
\cite{Starobinsky1980}, as the first inflationary model. In this
case the cosmic acceleration ends for $\alpha R^2<R$. Moreover,
Capozziello proposed an $f(R)$ model to obtain the cosmic
late-time acceleration \cite{Capozziello2002}, where he puts his
higher order gravity model in the category of modified matter
models. However in this paper, we concern about another model,
where $f(R)=R\pm\mu^{2(n+1)}/R^n$ (in this paper we consider $n=1$).
The model has been proposed in \cite{Capozziello2003a,
Capozziello2003b, Carroll2004, Nojiri2003}. For
$R^{n+1}>>\mu^{2(n+1)}$, we have $\frac{f(R)}{R}\rightarrow 1$, so the
$\mu$-dependent modification is vanished in this limit. However
for $R^{n+1}<<\mu^{2(n+1)}$, we get $\frac{f(R)}{R}\approx 1\pm\mu^{2(n+1)}$; then one can expect the modification
to the scalar gravity. The minus case however, appeared to be
encountering several shortcomings. For example the matter
instability \cite{Dolgov2003}, absence of the matter domination
era \cite{Amendola2007a, Amendola2007b} and inability to satisfy
local gravity constraints \cite{Chiba2003, Olmo2005a, Olmo2005b,
navarro2007, Chiba2007}. These problems stem from the fact that in
this model, $f''(R)<0$. However for $f(R)=R+\mu^{2(n+1)}/R^n$, we
have $f''(R)>0$ and it has been shown that in this case, the
problems are vanished and the theory becomes stable
\cite{Nojiri2003}, as well as some other models \cite{Nojiri2007,
Nojiri2008}. Furthermore it has been proved that this model
retains the matter domination era \cite{Sawicki2007}. In order to be in agreement with solar system experiments, in two interesting papers \cite{Saaidi2010, Saaidi2012} the authors obtain static
spherically symmetric solutions the case of $n=1$, i.e. for
$R\pm\mu^4/R$ theory of gravity, in both contexts of weak and
strong gravitational fields. In forthcoming sections, we consider
null-like geodesic congruences in the spacetimes described by the
positive part of mentioned solutions.

\section{Null Flows in the Weak Field Solution of $R+\mu^4/R$ Gravity}
The weak field static spherically symmetric solution for
$R+\mu^4/R$ gravity in solar system is ($c=1$)
\begin{equation}\label{3}
    \textrm{d}s^2=-A(r)\textrm{d}t^2+\frac{1}{B(r)}\textrm{d}r^2+r^2\textrm{d}\Omega^2,
\end{equation}
with \cite{Saaidi2010}
\begin{equation}\label{4}
    A(r)=1-\frac{2 M}{r}+\frac{3}{4} \alpha  (\mu  r)^{4/3},$$$$
    B(r)=1-\frac{2 M}{r}+\alpha  (\mu  r)^{4/3},
\end{equation}
in which $M$ is the Schwarzschild massive source and in the
additional terms to Schwarzschild  metric, $\alpha=(4/147)^{1/3}$.
Any freely falling particle in the gravitational field described
by a spacetime defined in (\ref{3}), must move on a geodesic,
obtained from the following geodesic equations:
\begin{equation}\label{5}
    {A' \dot r \dot t}+A\ddot t=0,
    $$$$
   B^2 \left(A \dot t^2-2 r \left(\dot \theta^2+\sin ^2(\theta ) \dot\phi^2\right)\right)-B' \dot r^2+2B\ddot r=0,
    $$$$
    r\ddot\theta-r\sin (\theta )\cos (\theta ) \dot\phi^2+{2 \dot\theta \dot
    r}=0,
    $$$$
    {2 \dot\phi \left(\dot r+r \cot (\theta )
    \dot\theta\right)}+r\ddot\phi=0.
\end{equation}
Form now on, prime stands for $\frac{\textrm{d}}{\textrm{d}r}$ and
dot for $\frac{\textrm{d}}{\textrm{d}\tau}$, with $\tau$ as the
trajectory parameter. Any congruence of curves, which are obtained
by integrating the above geodesic equations, form a flow of
integral curves in the spacetime. Moreover, if massless particles
are considered, we should also take into account the null
condition $g_{ab}\dot x^a\dot x^b=0$; .i.e.
\begin{equation}\label{6}
   -AB \dot t^2+{\dot r^2}+Br^2 \left(\dot \theta^2+\sin^2(\theta ) \dot \phi^2\right)=0.
\end{equation}
So the flow obeying (\ref{6}), is indeed a null flow, consisting
of for example light rays, or light flows. Now let us consider a
2-space, through which the light rays go in or go out. Either of
outgoing and ingoing flows are led by a tangential null vector
field, respectively $l^a$ and $n^a$. Therefore these vectors
satisfy $l^a l_a=n^a n_a=0$. Now consider a 2-space defined by the
metric \cite{Poisson2004}
\begin{equation}\label{7}
    h_{ab}=g_{ab}+l_a n_b + l_b n_a,
\end{equation}
orthogonal to both of outgoing and ingoing vectors, i.e. $h_{ab}
l^a=h_{ab} n^a=0$. Hence in order to satisfy this, an
additional condition $l^a n_a=-1$ is mandatory, so that ${h^a}_a=2$.

Let us define a tensor
\begin{equation}\label{7'}
X_{ab}=\nabla_a l_a,
\end{equation}
orthogonal to the null congruence, i.e. $X_{ab} l^a=X_{ab} l^b=0$. One can decompose $X_{ab}$ to symmetric 
and anti-symmetric parts
\begin{equation}\label{8'}
X_{ab}=\theta_{ab}+\omega_{ab},
\end{equation}
where the symmetric part $\theta_{ab}$ itself, can be decomposed to trace and traceless parts
\begin{equation}\label{9'}
\theta_{ab}=\frac{1}{2}\,\Theta\,h_{ab}+\sigma_{ab}.
\end{equation}
Inclusion in (\ref{8'}) and taking the trace ${X_a}^a$ yields
\begin{equation}\label{10'}
{X_a}^a=\Theta=\nabla_a l^a.
\end{equation}
The scalar expansion $\Theta$ is the fractional rate of change of the congruence, per unite affine parameter, 
in the transverse cross-sectional area described by the 2-metric $h_{ab}$. Moreover, the traceless part of (\ref{9'})
is defined as
\begin{equation}\label{14}
    \sigma_{ab}=\nabla_{(a}l_{b)}-\frac{\Theta}{2}h_{ab},
\end{equation}
which is the symmetric shear. Also the anti-symmetric part of (\ref{8'}) is the anti-symmetric vorticity
\begin{equation}\label{15}
    \omega_{ab}=\nabla_{[a}l_{b]}.
\end{equation}
These kinematical characteristics constitute the outstanding Raychaudhuri equation \cite{Poisson2004, Raychaudhuri1955}
\begin{equation}\label{12}
   \dot\Theta=-\frac{1}{2}\Theta-\sigma^2+\omega^2-R_{ab} l^a l^b,
\end{equation}
with $\sigma^2=\sigma_{ab}\sigma^{ab}$ and $\omega^2=\omega_{ab}\omega^{ab}$ in which $\sigma^{ab}=g^{ae} g^{bf} \sigma_{ef}$ and 
$\omega^{ab}=g^{ae} g^{bf} \omega_{ef}$ are obtained with respect to the background spacetime metric.
 The Raychaudhuri equation (\ref{12}), has appeared to be a quite useful mathematical tool to
investigate the focusing theorem and the concept of singularity
\cite{Penrose1965, Hawking1970, Hawking1973}.
To go any further, we confine our flow to move only in equatorial plane
($\theta=\pi/2$), so any tangential outgoing vector field $l^a$ in
a sapcetime like (\ref{3}) and in equatorial plane can be defined
as
\begin{equation}\label{8}
    l^a=\left(\dot t, \dot r, 0, \dot \phi\right).
\end{equation}

\subsection{Pure Radial Flow}
In this case we consider $\phi=$const. So integrating the first in
(\ref{5}) results in
\begin{equation}\label{9}
   \dot t=\frac{E}{A},
\end{equation}
where $E$, based on the definition $E=-g_{00}\dot t$
\cite{Misner1973}, is the positive energy of moving particles and
is indeed a constant of motion. Accordingly, and using the null
condition (\ref{6}), we get
\begin{equation}\label{10}
    l^a=E\left(\frac{1}{A},-\sqrt{\frac{B}{A}},0,0\right).
\end{equation}
The condition $l^a n_a=-1$ for the outgoing null vector $n^a$,
provides
\begin{equation}\label{11}
    n^a=\frac{1}{2E}\left(1, \sqrt{AB}, 0, 0\right).
\end{equation}
Now let us introduce three important kinematical characteristics
of affinely parameterized geodesic flows, like the ones defined by
(\ref{10}) and (\ref{11}). In (\ref{12}) we
define
\begin{equation}\label{13}
\Theta_l=\nabla_a l^a,
$$$$
\Theta_n=\nabla_a n^a,
\end{equation}
to be the scalar expansion, respectively of affinely parameterized
ingoing and outgoing flows. These parameters are indeed in
the 2-space $h_{ab}$, however if we are interested in what it
really is, we should examine both orthogonally ingoing and
outgoing flows through this 2-space, while they are propagated in
the gravitational field. The expansions (\ref{13}) for the weak
field values of $R+\mu^4/R$ gravity, given in (\ref{4}) and
corresponding tangential vectors (\ref{10}) and (\ref{11}) are
\begin{equation}\label{16}
    \Theta_l=-\frac{4E}{r}\sqrt{\frac{2 M-r \left(\alpha  (\mu  r)^{4/3}+1\right)}{8 M-r \left(3 \alpha  (\mu
    r)^{4/3}+4\right)}},
    $$$$
    \Theta_n=\frac{r \left(5 \alpha  (\mu  r)^{4/3}+4\right)-4 M}{2 E r^2}\sqrt{\frac{ 2 M-r \left(\alpha  (\mu  r)^{4/3}+1\right)}{8 M-r \left(3 \alpha  (\mu
    r)^{4/3}+4\right)}}.
\end{equation}
Here, we bring a discussion on the types of 2-surfaces which are
directly extrapolated from the values in (\ref{16}), based on the
information in \cite{Booth2005, Nielsen2009, Gourghoulhon2008,
Ashtekar2005,  Booth2006} (also for a very good review see
\cite{Faraoni2013}).

\begin{itemize}
\item{For $h_{ab}$ to be a normal surface metric (like a 2-sphere
in Minkowski spacetime), one must have $\Theta_l>0$ and
$\Theta_n<0$.}

\item{A trapped surface is obtained if \cite{Penrose1965}
$\Theta_l<0$ and $\Theta_n<0$. Accordingly, both future directed
ingoing and outgoing flows are converged at a singularity. Therefore, trapped surfaces correspond 
to black hole regions.}

\item{For the outgoing flow $l^a$, if $\Theta_l=0$ and
$\Theta_n<0$ for the ingoing flow, a marginal surface is
available.}

\item{If $\Theta_l\Theta_n<0$, then $h_{ab}$ is an un-trapped
surface.}

\item{Finally, an anti-trapped surface is obtained, when
$\Theta_l>0$ and $\Theta_n>0$, which leads to diverging future
directed outgoing and ingoing flows. This mean that all data will scape from the surface; representing a white hole region \cite{Carroll20041}.}

\end{itemize}

For the outgoing flow expansion (\ref{16}) we always have
$\Theta_l<0$, except for the value
\begin{equation}\label{17}
    \mu=\pm\left(\frac{2 M-r}{\alpha}\right)^{3/4} \frac{1}{r^{7/4}},
\end{equation}
where $\Theta_l=0$. Therefore we do not possess a normal surface here. Moreover, the
ingoing flow expansion $\Theta_n$ in (\ref{16}), vanishes for
\begin{equation}\label{18}
    \mu=\pm\left(\frac{2 M-r}{\alpha}\right)^{3/4}
    \frac{1}{r^{7/4}}
    \,\,\,\,\textrm{and}\,\,\,\,=\pm\left(\frac{ M-r}{5
    \alpha}\right)^{3/4}
    \frac{2 \sqrt{2}}{r^{7/4}}.
\end{equation}
On the other hand, we can not expect a marginal surface, since for the value in (\ref{17}) for $\Theta_l$, we have also
$\Theta_n=0$. However, adjacent to Schwarzschild radius
$r\approx2M$, we have
\begin{equation}\label{19}
   \Theta_n\approx \frac{5 \alpha  \mu  M \sqrt[3]{2\mu  M}+1}{2\sqrt{3}
   E M},
\end{equation}
which is always positive. So according to the fact that always
$\Theta_l<0$, then $\Theta_l \Theta_n<0$, which corresponds to an
un-trapped surface. Figure 1 shows the behaviour of $\Theta_n$ with
respect to $\mu$, near the Schwarzschild radius.

\begin{figure}
 \center{ \includegraphics[width=8cm]{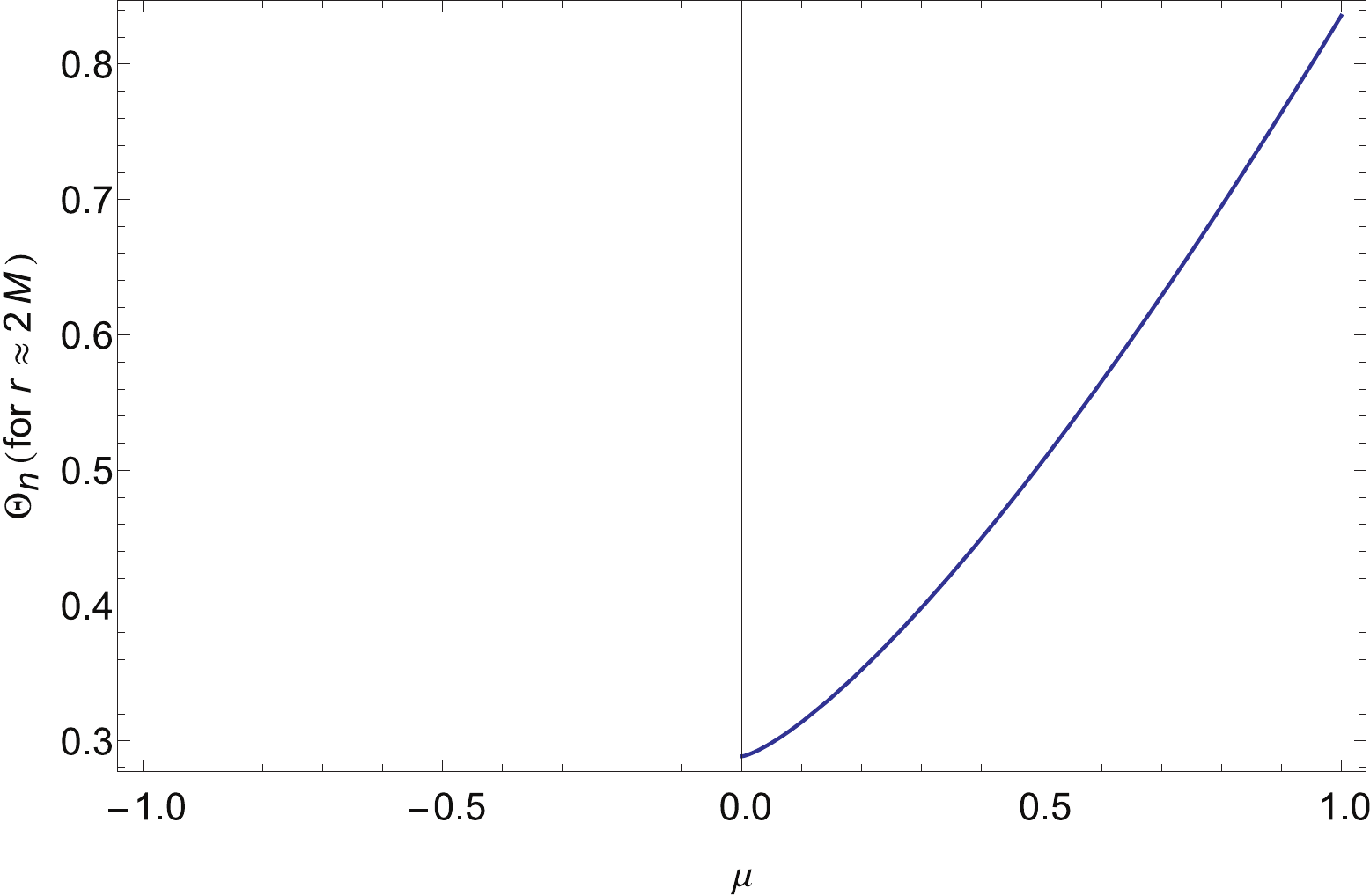}\\
  \caption{\small{The behaviour of $\Theta_n$ with respect to $\mu$, for $r\approx2M$ for radial flows. The plotting has been done for $E=1$, and the unit value along the $\mu$ axis is $M$. Obviously for any values of $\mu$, always $\Theta_n>0$.}}}
\end{figure}

\subsection{Radial-Rotational Flow}
In this case, the geodesic equations become
\begin{equation}\label{20}
    \dot t=\frac{E}{A},
    $$$$
    \dot \phi=\frac{L}{r^2},
\end{equation}
in which $L=g_{33} \dot\phi$, is the proper angular momentum of
moving particles, and is another constant of motion
\cite{Misner1973}. So applying the null condition (\ref{6}), the
tangential vector $l^a$ for outgoing flow is obtained as
\begin{equation}\label{21}
    l^a=\left(\frac{E}{A},\sqrt{B}
    \sqrt{\frac{E^2}{A}-\frac{L^2}{r^2}},0,\frac{L}{r^2}\right).
\end{equation}
Accordingly, the condition $l^an_a=-1$, may result in the
following for the null ingoing vector:
\begin{equation}\label{22}
    n^a=\frac{r^2}{L^2 A}\left(E (L+1)-\sqrt{(2 L+1) A}
    \sqrt{\frac{E^2}{A}-\frac{L^2}{r^2}}, \,\,\,\,(L+1) A \sqrt{B} \sqrt{\frac{E^2}{A}-\frac{L^2}{r^2}}-\sqrt{E^2 (2 L+1) A
    B},\,\,\,\, \,\,\,\,0,\,\,\,\, 1 \right).
\end{equation}
Therefore the outgoing expansion would be
\begin{equation}\label{23}
   \Theta_l=\frac{\sqrt{-\frac{2 M}{r}+\alpha  (\mu  r)^{4/3}+1} \left[8 E^2 r^3+L^2 \left(4 M-r \left(5 \alpha  (\mu  r)^{4/3}+4\right)\right)\right]}{r^2 \sqrt{r \left(3 \alpha  (\mu  r)^{4/3}+4\right)-8 M} \sqrt{4 E^2 r^3+L^2 \left(8 M-r \left(3 \alpha  (\mu  r)^{4/3}+4\right)\right)}}
\end{equation}
The result for $\Theta_n$ becomes rather complicated, however once
again we can plot the Schwarzschild limit of both $\Theta_l$ and
$\Theta_n$. Figure 2 shows these values which has been plotted in
terms of $\mu$, for different values of $E$ and $L$.

\begin{figure}
\center{  \includegraphics[width=5cm]{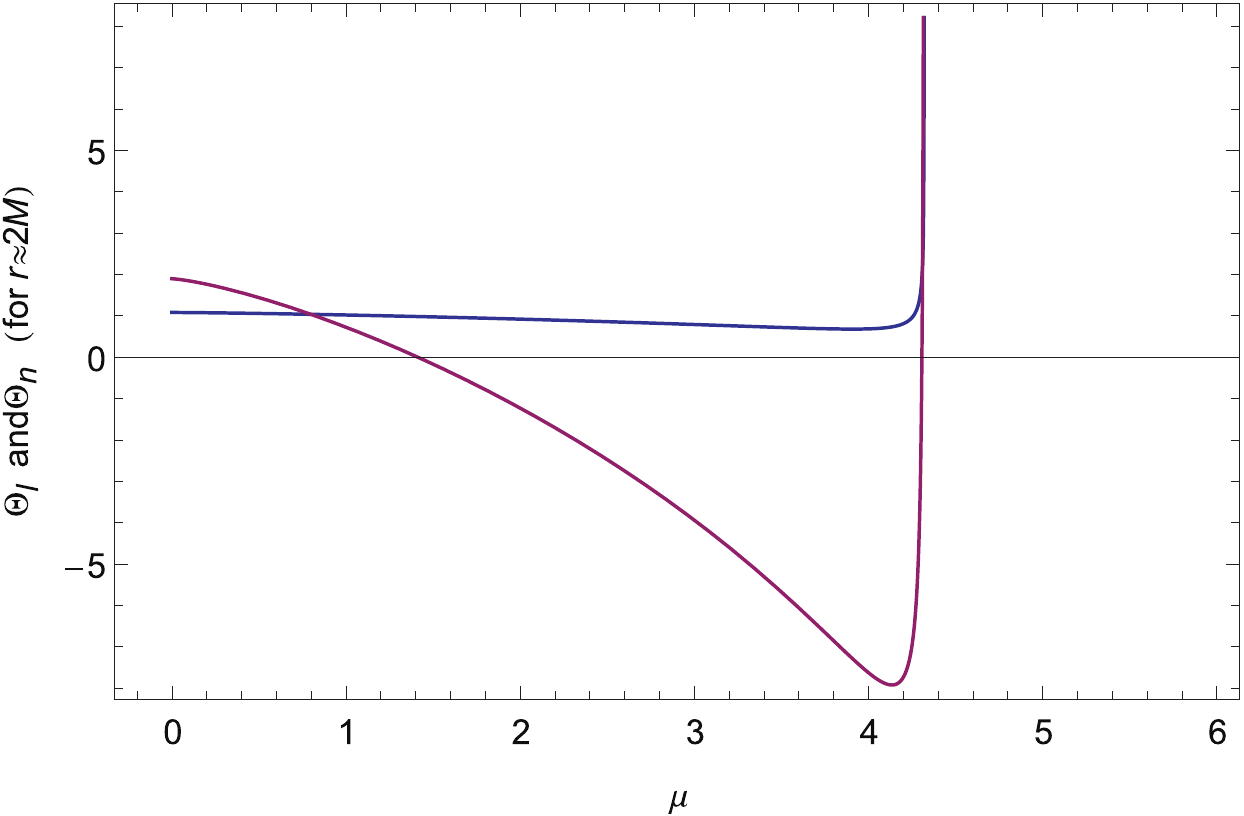}~a) \hfil
\includegraphics[width=5cm]{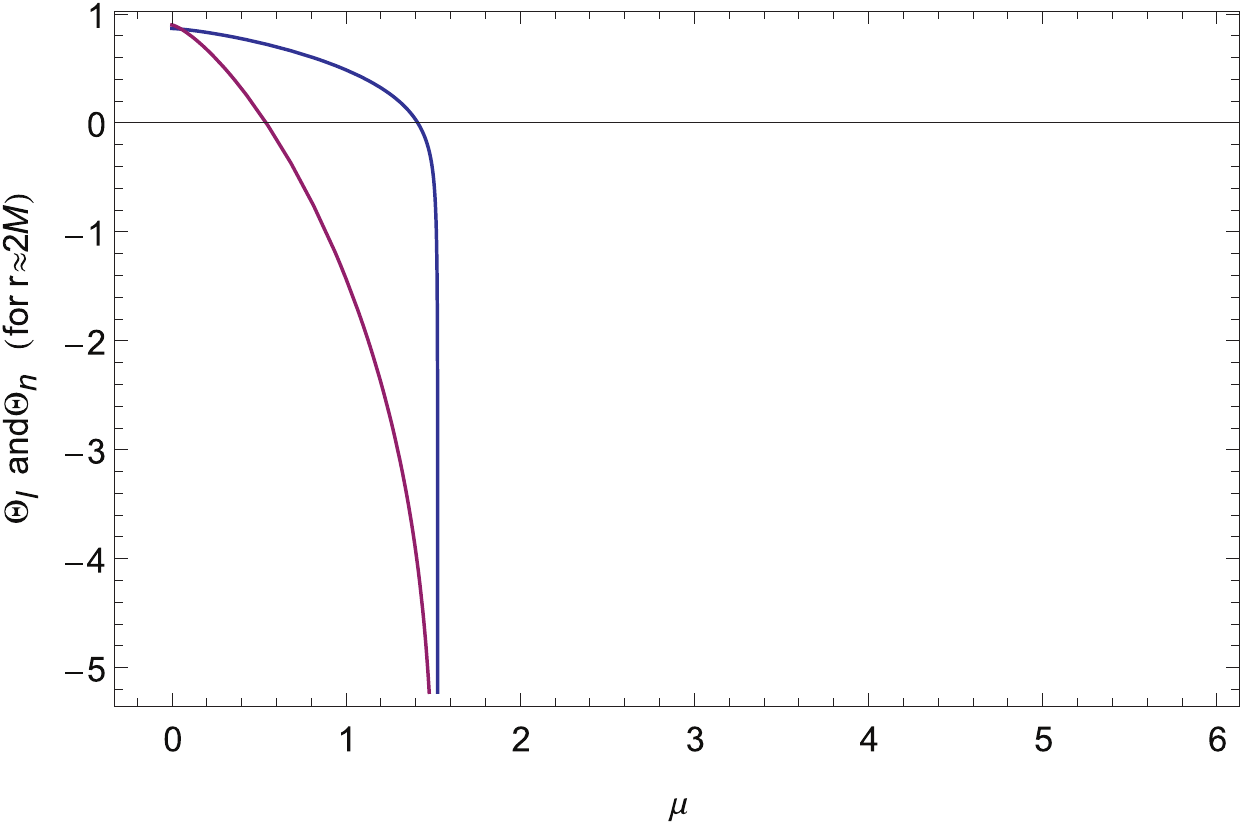}~b)
\hfil
\includegraphics[width=5cm]{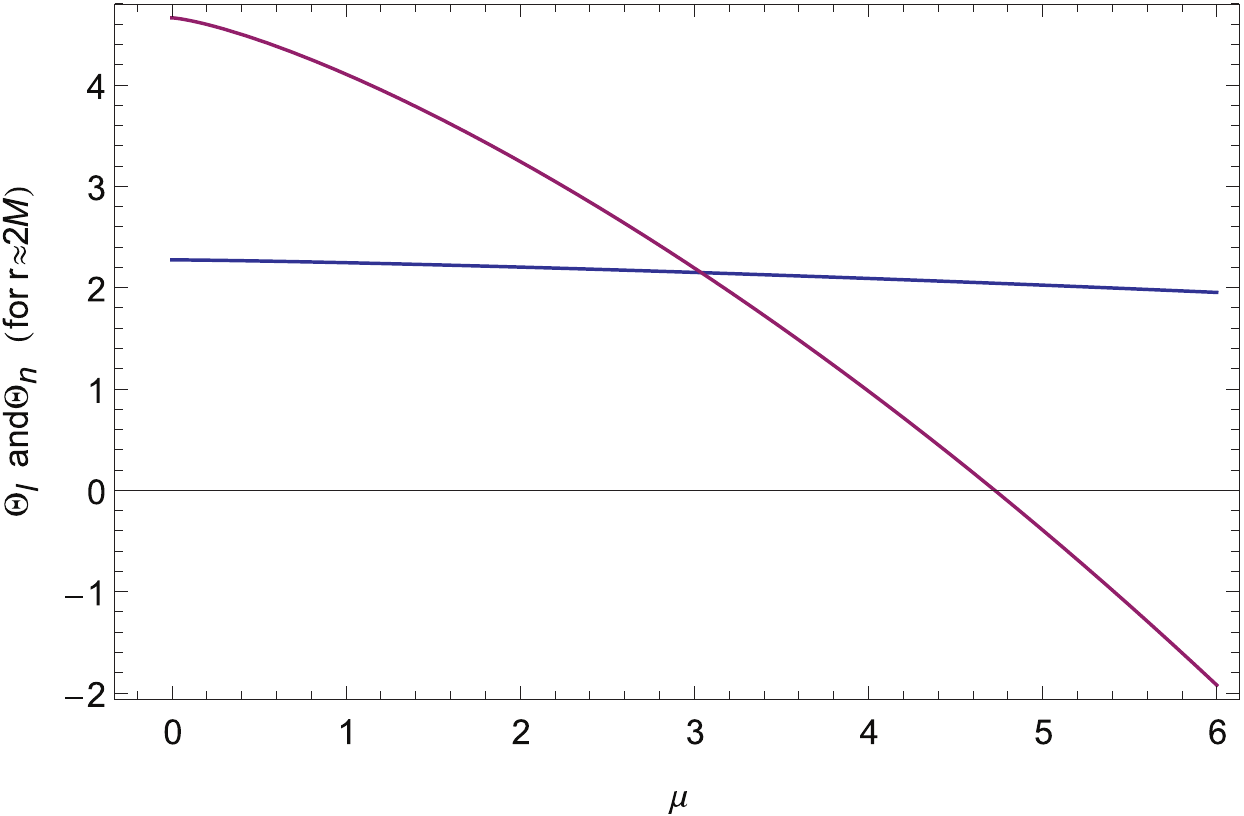}~c)
\caption{The behaviour of $\Theta_l$ (blue line) and $\Theta_n$
with respect to $\mu$, for $r\approx2M$ for radial-rotational congruence. The
plotting has been done for (\textbf{a}) $E=1$ and $L=1$,
(\textbf{b}) $E=1$ and $L=2$ and (\textbf{c}) $E=2$ and $L=1$. the
unit value along the $\mu$ axis is $M$.}}
\end{figure}

\begin{figure}
\center{  \includegraphics[width=5cm]{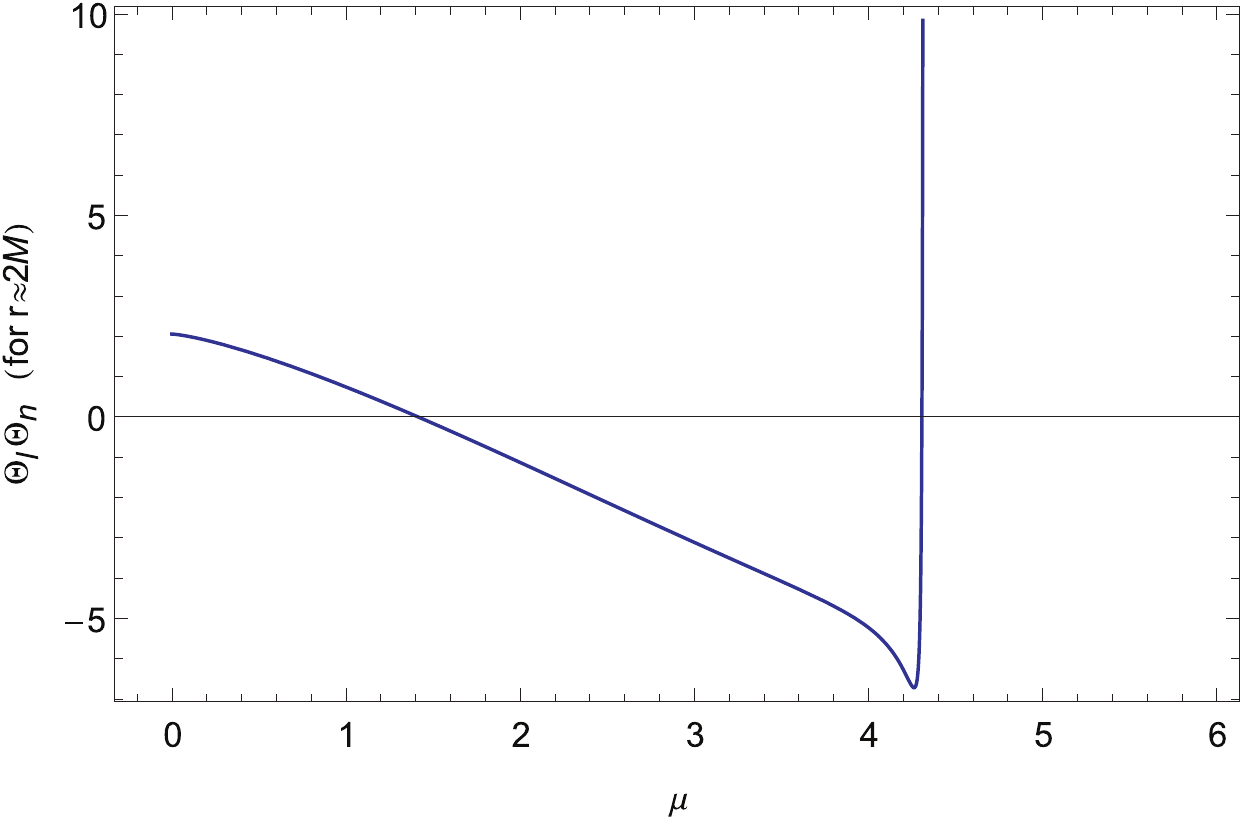}~a) \hfil
\includegraphics[width=5cm]{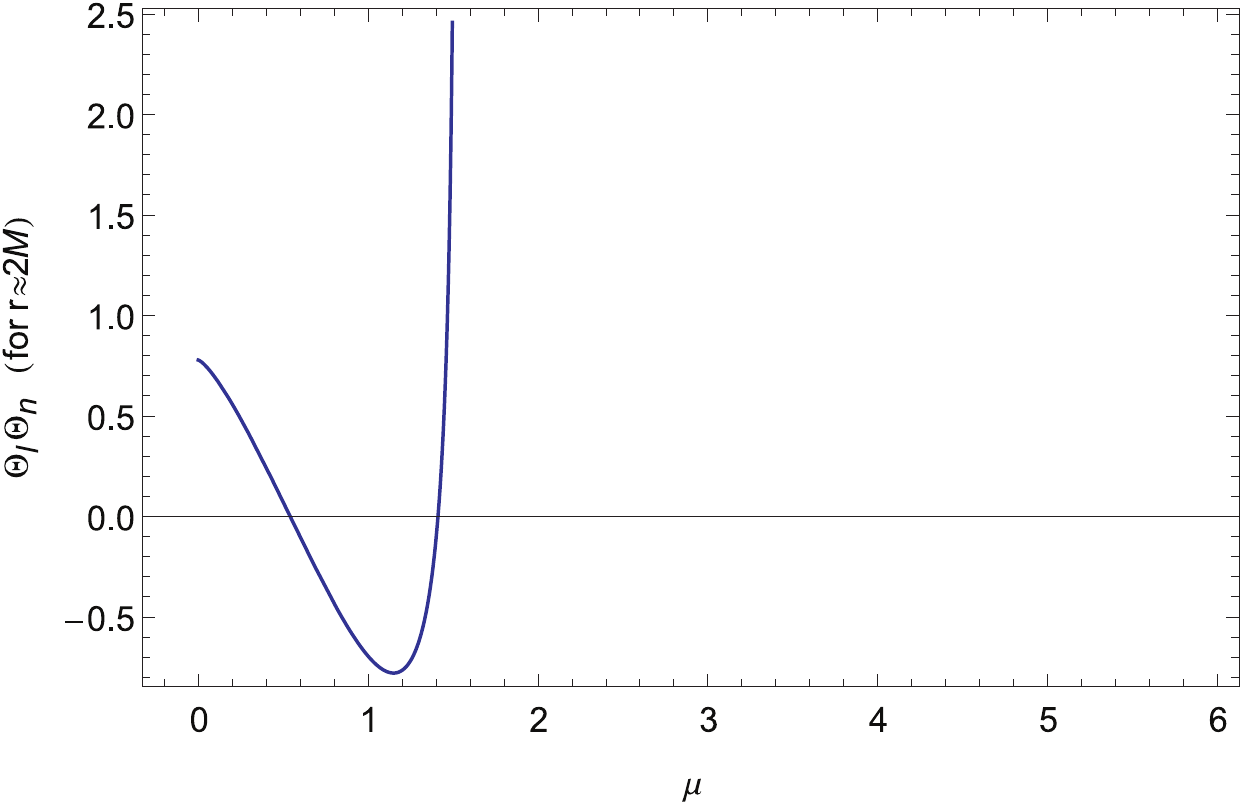}~b)
\hfil
\includegraphics[width=5cm]{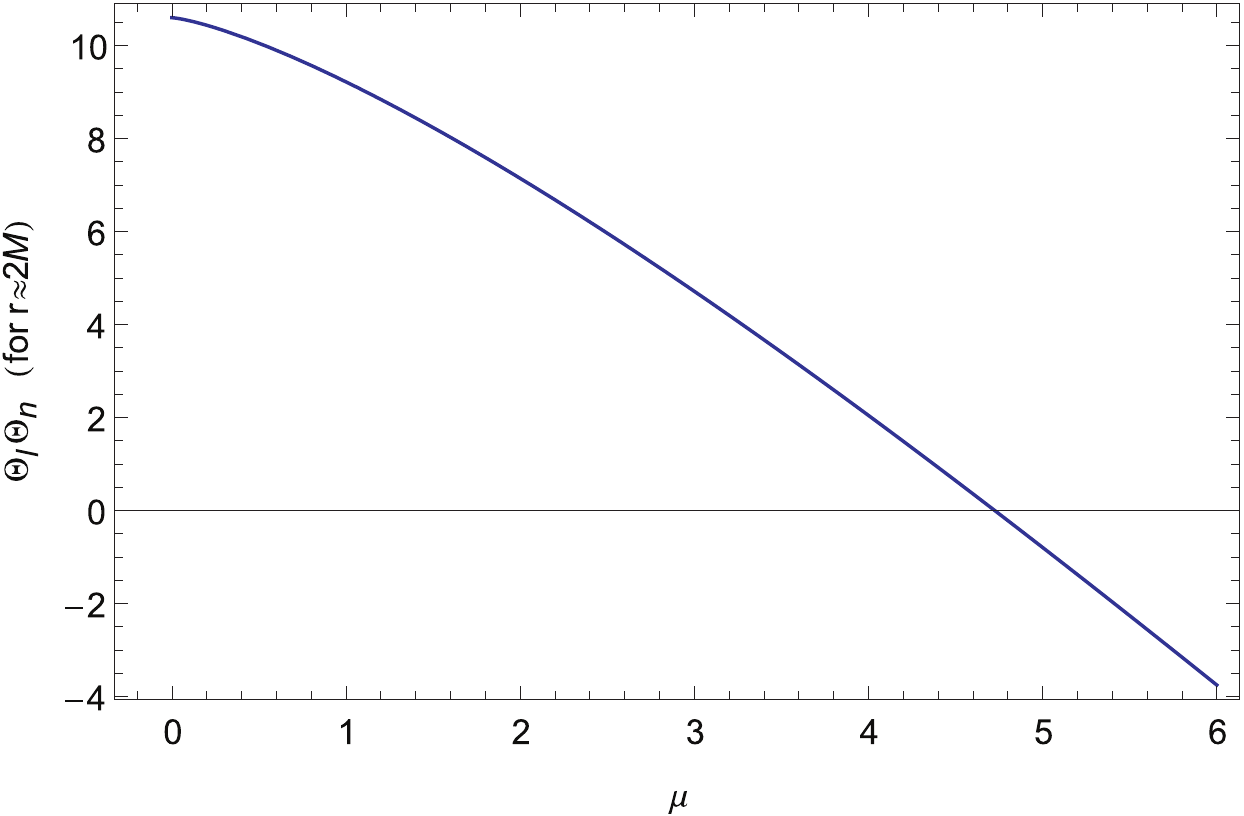}~c)
\caption{The behaviour of $\Theta_l\Theta_n$ with respect to $\mu$,
for $r\approx2M$ for radial-rotational. The plotting has been done
for (\textbf{a}) $E=1$ and $L=1$, (\textbf{b}) $E=1$ and $L=2$ and
(\textbf{c}) $E=2$ and $L=1$. the unit value along the $\mu$ axis
is $M$. As it is seen, these may include un-trapped, trapped,
anti-trapped and marginal surfaces.}}
\end{figure}
The expansions do not have real values for $\mu<0$. One can see that
at some points we encounter both $\Theta_l<0$ and $\Theta_n<0$
which corresponds to trapped surfaces for $h_{ab}$, or black hole
event horizons. However according to figure 2, one can see that
the values of $\Theta_l \Theta_n$ are somewhere negative and
somewhere positive. This can provide that we have un-trapped
surfaces as well as trapped and anti-trapped ones. Moreover, for
$\Theta_l \Theta_n=0$ in figure 3, marginal surfaces may be
available.

\section{Strong Field Counterpart}

Given in \cite{Saaidi2012} and corresponding to (\ref{3}), the
strong field static solution to $R+\mu^4/R$ gravity is
\begin{equation}\label{24}
    A(r)=-\frac{1}{8} 3 \sqrt[3]{\frac{4}{3}} (2 \mu  M)^{4/3} \left(\frac{r}{2 M}-1\right)^{2/3}-\frac{2
    M}{r}+1,$$$$
    B(r)=\frac{1}{\frac{1}{8} \sqrt[3]{\frac{4}{3}} (2 \mu  M)^{4/3} \left(\frac{r}{2 M}-1\right)^{2/3}-\frac{2
    M}{r}+1}.
\end{equation}
Accordingly, the value of $\Theta_l\Theta_n$ in the Schwarzschild
limit (with $M=1$) becomes
\begin{equation}\label{25}
    \Theta_l\Theta_n\approx -\frac{2 \left(6^{2/3} \mu ^{8/3}-2 \sqrt[3]{2} \mu ^{4/3} \sqrt[3]{r-2}+4 \sqrt[3]{3} (r-2)^{2/3}\right) \left(6^{2/3} \mu ^{8/3}-2 \sqrt[3]{2} \mu ^{4/3} \sqrt[3]{r-2}+3 \sqrt[3]{3} \mu ^4 (r-2)^{2/3}+4 \sqrt[3]{3} (r-2)^{2/3}\right)}{3\ 3^{2/3} \mu ^8
    (r-2)^{4/3}},
\end{equation}
which does not depend on the energy $E$. One can plot while $r$ is still
changing in the desired domain (see figure 4). Apparently for all
values of $\mu$ near $r=2M$, always $\Theta_l\Theta_n<0$. So we
possess un-trapped surfaces.

\begin{figure}
\center{  \includegraphics[width=8cm]{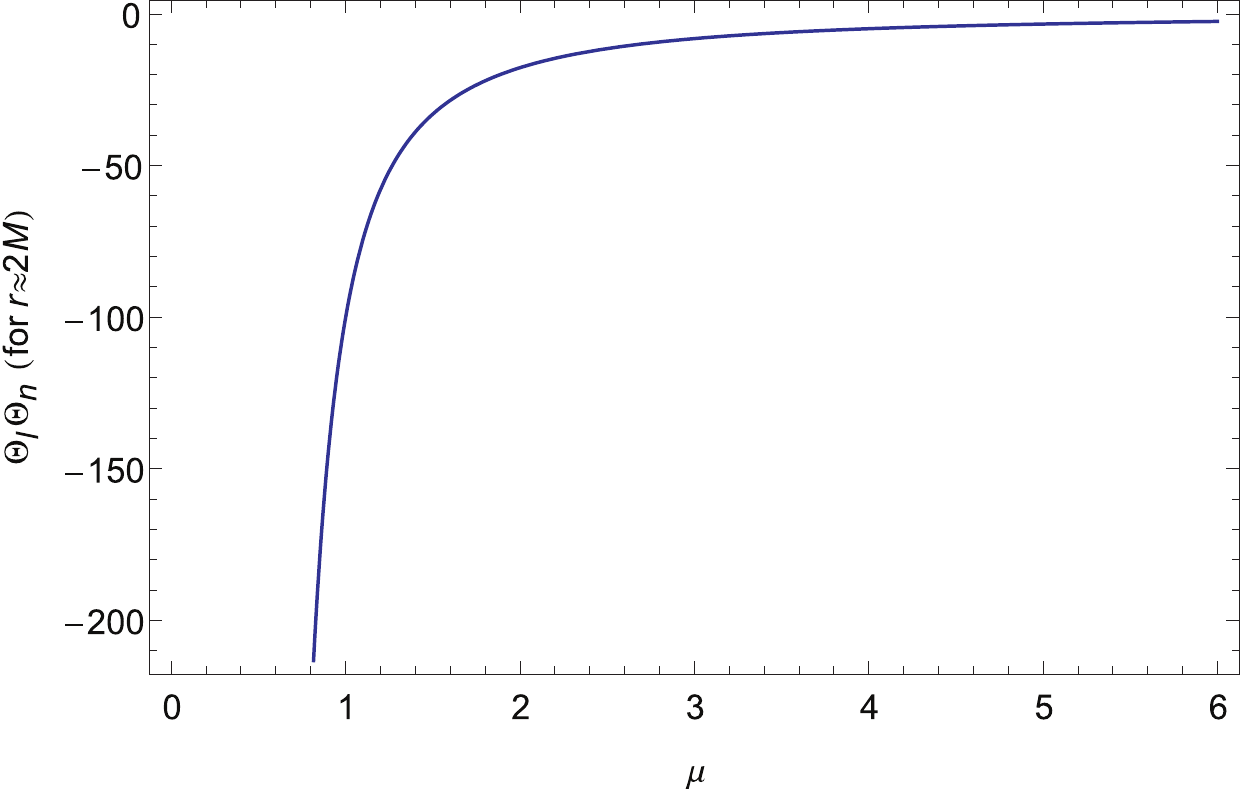} \caption{The behaviour of
$\Theta_l\Theta_n$ with respect to $\mu$, for $r\approx2M$ in
strong fields for radial flows. The plotting has been done $r=2.1$
the unit value along the $\mu$ axis is $M$.}}
\end{figure}

If radial-rotational flows are considered, the values of
$\Theta_l\Theta_n$ form a curve like figure 4 (see figure 5),
however for very larger values. It turns out that these values are
all negative, therefore once again we encounter un-trapped
surfaces.

\begin{figure}
\center{  \includegraphics[width=8cm]{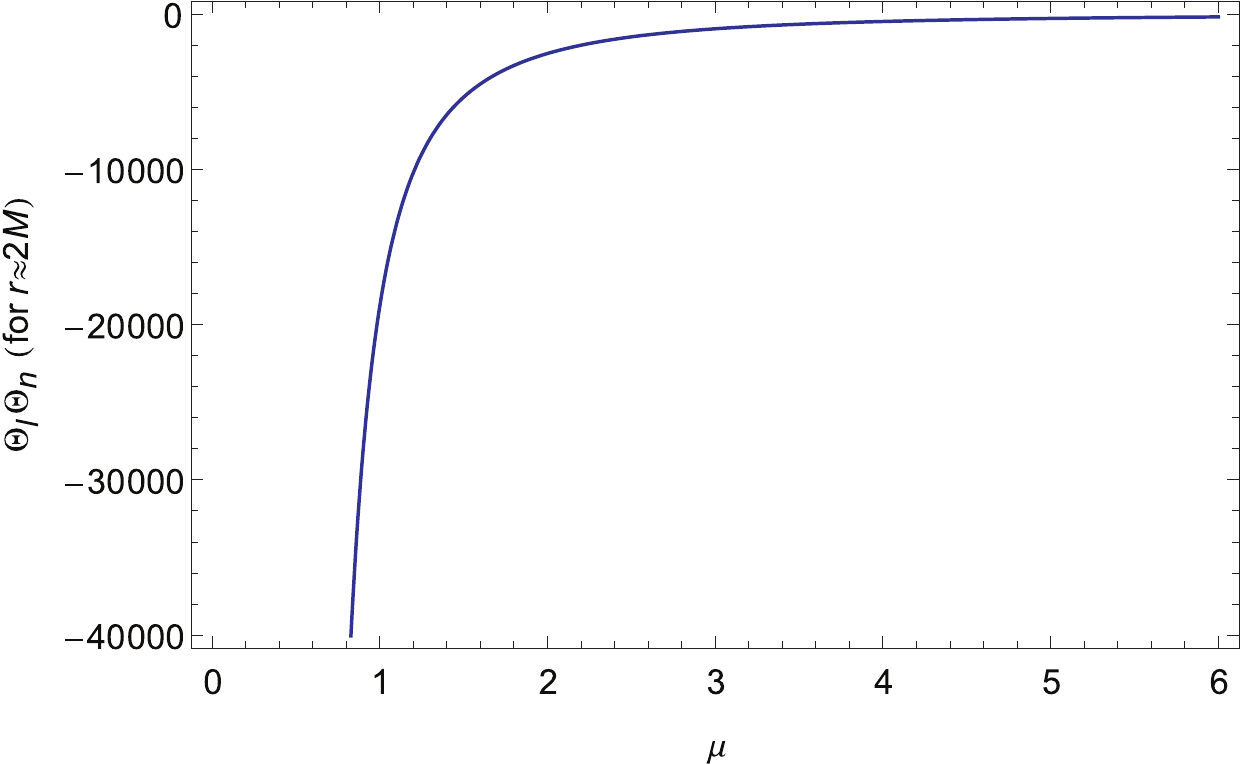} \caption{The behaviour of
$\Theta_l\Theta_n$ with respect to $\mu$, for $r\approx2M$ in
strong fields for radial-rotational flows. The plotting has been
done for $E=1$, $L=1$ and $r=2.1$ the unit value along the $\mu$
axis is $M$.}}
\end{figure}

\subsection{A Note on the Initial Values for Radial-Rotational Flows}
If the particles are commencing motion (at $\tau=0$) form a
initial radial point $r_0$, where $r_0=r(\tau)|_{\tau=0}$, and
with temporal initial velocity $u_t=\dot t|_{\tau=0}$, radial
initial velocity $u_r=\dot r|_{\tau=0}$ and initial angular
velocity $\omega_0=\dot\phi|_{\tau=0}$, then the values of energy
$E$ and angular momentum $L$ in (\ref{20}) can be rewritten as
\begin{equation}\label{20-1}
    E=u_t A_0,$$$$
    L=\omega_0 {r_0}^2,
\end{equation}
where $A_0=A(r_0)$. Now if we consider the particles to be
released from a circular orbit of radius $r_0$, then we can ignore
$u_r$, since by the time the particles are released ($\tau=0$),
they are at a constant position $r=r_0$ on a circle. Hence, the
corresponding null condition (\ref{6}) at $\tau=0$ gives
\begin{equation}\label{20-2}
{\omega_0}^2=\frac{A_0}{{r_0}^2}u_t^2.
\end{equation}
Also from (\ref{21}) and (\ref{20-1}) we have
\begin{equation}\label{20-3}
    \sqrt{B_0}\sqrt{\frac{{A_0}^2u_t^2}{A_0}-(r_0\omega_0)^2}=0.
\end{equation}
Therefore using (\ref{20-2}), one can obtain
\begin{equation}\label{20-4}
   u_t= \pm\frac{r_0}{\sqrt{A_0}}.
\end{equation}
So both constants of motion, can be expressed in terms of the
initial radius of release $r_0$.

\section{Conclusion}
In this paper we investigated the 2-dimensional cross-sectional surfaces of exterior geometries of $R+\mu^4/R$
theory of gravity, through which the null congruence of geodesic
integral curves (null flows) can pass. We pursued both weak field
and strong field limits. Our method was based on inspecting the
expansion of the outgoing and ingoing flows, in which the signs of
these expansions were crucial to the type of the mentioned
surface. According to different types of motion, we obtained
different types of surfaces, including black hole horizons
(trapped surfaces). However, this type of surface was rather rare,
since most of the types included un-trapped surfaces. Foe purely radial congruence in weak 
field limit, we discovered that no matter $\mu$ is positive or negative, we do not encounter a black hole, since always 
$\Theta_l \Theta_n<0$. Once radial-rotational flows are assumed, then one can observe that in order to have real values, $\mu>0$ is mandatory. Regarding its evolution, these positive values can affect $\Theta_l \Theta_n$ in a way that in terms of $\mu$, the surface may evolve from a trapped (black hole) or anti-trapped (white hole) surface, to a marginal and eventually to an un-trapped surface. This can be seen in all three cases of figure 3. However,for evolving surfaces in the strong field limit, in both cases of radial and radial-rotational congruence and regardless of numerical values, $\Theta_l \Theta_n$, is always negative. This means that the cosmological term $\mu$ imposes a rather reach effect on the gravitational attraction, enforcing $h_{ab}$ to define an un-trapped surface. In this case, the extra term $\mu$, as a cosmological background on Schwarzschild geometry, causes some sort of repelling force, which cancels out the possibility of a gravitational collapse. So, it seems that to find possible $f(R)$ black holes in $R+\mu^4/R$ gravity, one must take only its weak field limit geometry.

\subsubsection*{\textbf{Acknowledgements} \,\,\,\, \textmd{We would like to thank the referee for useful comments, which helped us improve the presentation of the paper}.}

\end{document}